\RequirePackage{ifpdf}
\ifpdf
\else
\PassOptionsToPackage{dvipdfmx}{graphicx}
\PassOptionsToPackage{dvipdfmx}{hyperref}
\fi

\documentclass[12pt,a4paper]{article}
\usepackage{jheppub}

\allowdisplaybreaks[1]
\usepackage{amsmath,bm}
\usepackage{subfigure}

\title{
Resonating AdS soliton
}

\author[1]{Markus Garbiso,}
\author[2]{Takaaki Ishii,}
\author[3]{and Keiju Murata}

\affiliation[1]{Department of Physics and Astronomy, University of Alabama, Tuscaloosa, AL 35487, USA}
\affiliation[2]{Department of Physics, Kyoto University, Kyoto 606-8502, Japan}
\affiliation[3]{Department of Physics, College of Humanities and Sciences, Nihon University, Tokyo 156-8550, Japan}
\emailAdd{magarbiso@crimson.ua.edu}
\emailAdd{ishiitk@gauge.scphys.kyoto-u.ac.jp}
\emailAdd{murata.keiju@nihon-u.ac.jp}

\abstract{%
The AdS soliton is a nonsingular spacetime that has a flat conformal boundary with a compact $S^1$ direction. We find a horizonless cohomogeneity-1 metric that describes nonlinear gravitational oscillations of the AdS soliton in five dimensions. We call this spacetime {\it the resonating AdS soliton}. This solution is obtained as the nonlinear extension of  normal modes of the AdS soliton dual to spin-2 glueball excitations. The boundary energy momentum tensor of the resonating AdS soliton has time periodic components, and it is interpreted as a coherently excited state in the dual field theory. Physical quantities of the resonating AdS soliton are multivalued at a fixed energy, suggesting a transition between different frequency solutions. The energy of the resonating AdS soliton is higher than that of the undeformed AdS soliton, in accordance with the positive energy conjecture proposed by Horowitz and Myers.
}

\preprint{KUNS-2820}

\begin{document}
\maketitle

\section{Introduction}
\label{sec:intro}

The AdS soliton~\cite{Horowitz:1998ha} is a globally non-singular solution that has negative energy relative to  the pure AdS space. It is an asymptotically locally AdS spacetime and has a flat conformal boundary with a compact $S^1$ circle. It has been conjectured that the AdS soliton is the unique lowest energy solution among the spacetimes with the same asymptotic structure \cite{Horowitz:1998ha,Constable:1999gb}. The uniqueness theorem for the AdS soliton supports the conjecture \cite{Galloway:2001uv,Galloway:2002ai}, but its full proof is yet to be given \cite{Woolgar:2016axs}.\footnote{Recently, results supporting the Horowitz-Myers conjecture were given also in \cite{Barzegar:2019vaj}.}

In the AdS/CFT duality \cite{Maldacena:1997re,Gubser:1998bc,Witten:1998qj}, the AdS soliton has been considered with respect to the gravity dual of the confining phase of pure Yang-Mills theory. In the dual field theory, an anti-periodic boundary condition for fermions along the $S^1$ breaks supersymmetry, and so pure Yang-Mills theory can be described \cite{Witten:1998zw}. In the bulk, the spacetime of the AdS soliton is smoothly capped at a finite distance where the $S^1$ shrinks, and gapped normal modes, corresponding to the confined spectrum, are obtained.
At finite temperature, there is a Hawking-Page transition \cite{Hawking:1982dh} between an AdS soliton and a black brane \cite{Surya:2001vj}, and this is interpreted as the confinement/deconfinement transition~\cite{Witten:1998zw}.\footnote{There is another interesting aspect of the AdS soliton; the dynamics of probe classical strings in the geometry are chaotic \cite{Basu:2011dg,Ishii:2016rlk} in contrast to their integrability in pure AdS space.}

In this paper, we will study time-periodic states in the confined phase of the gauge theory by using its gravity dual.
Understanding dynamical phenomena is one of the most challenging problems of strongly coupled quantum many body systems. 
The gauge/gravity duality helps us to understanding such problems.
The normal modes that the AdS soliton admits are expected to have a nonlinear extension.
In particular, we will focus on the spin-2 perturbation of the AdS soliton \cite{Constable:1999gb,Brower:1999nj} and show that its nonlinear extension leads to a non-stationary time periodic solution.
We will call such solutions as {\it the resonating AdS solitons}.

In particular, we will consider five-dimensional solutions with cohomogeneity-1 metrics.
In asymptotically global AdS space, similar setups for non-stationary solutions have been considered. Solutions describing nonlinear gravitational oscillations are called geons~\cite{Dias:2011ss,Horowitz:2014hja,Martinon:2017uyo,Fodor:2017spc}.
Analogously, there are boson stars, characterized by oscillating scalar fields \cite{Dias:2011at,Liebling:2012fv,Maliborski:2013jca,Buchel:2013uba,Fodor:2015eia,Choptuik:2017cyd,Choptuik:2018ptp}.
These solutions are non-stationary but time periodic. 
Geons constructed in four dimensions only have a single helical Killing vector~\cite{Horowitz:2014hja}. 
But recently, by going to five dimensions, geons whose metric is cohomogeneity-1 (i.e. the metric functions only depend on a single variable) have been constructed \cite{Ishii:2018oms,Ishii:2019wfs}. 
By applying techniques in Refs.~\cite{Ishii:2018oms,Ishii:2019wfs}, we construct the resonating AdS soliton with cohomogeneity-1 metric in five dimensions.
There has been a perturbative construction \cite{Hartnett:2012np}, but we are able to find non-trivial nonlinear solutions.

Gravitational dynamics in AdS space of general relativity is theoretically interesting.
Because of the timelike conformal boundary of the asymptotically AdS space, non-trivial dynamics can be realized.
One example is the weakly turbulent instability of global AdS space~\cite{Bizon:2011gg}: pure AdS is nonlinearly unstable against arbitrarily small perturbations, and a black hole is formed for a wide class of initial data. 
Similar phenomena has been observed in the AdS soliton background although the presence of a finite threshold of amplitude for the black hole formation is a significant difference~\cite{Craps:2015upq,Myers:2017sxr}.
Below the threshold, the spacetime oscillates forever and is time periodic. Our resonating AdS soliton might be one simple realization of such a phase.

The normal modes of the AdS space are connected to the onset of the superradiant instability of rotating black holes \cite{Hawking:1999dp,Reall:2002bh,Cardoso:2004hs,Kunduri:2006qa,Cardoso:2006wa,Murata:2008xr,Kodama:2009rq,Dias:2011at,Dias:2013sdc,Cardoso:2013pza}. 
Indeed, gravitational geons correspond to the smooth horizonless limit of black resonators \cite{Dias:2015rxy,Niehoff:2015oga,Ishii:2018oms,Ishii:2019wfs}.\footnote{Scalar field normal modes and superradiant instabilities also lead to analogous boson stars and hairy black hole solutions with similar dynamics these gravitational solutions \cite{Reall:2002bh,Cardoso:2004hs,Kunduri:2006qa,Cardoso:2006wa,Murata:2008xr,Kodama:2009rq,Dias:2011at,Dias:2013sdc,Cardoso:2013pza}.}
A black resonator is a dynamical spacetime. Therefore, it will be prone to further superradiant instabilities \cite{Green:2015kur}.
Stability analysis has been carried out for five-dimensional cohomogeneity-1 solutions \cite{Ishii:2020muv}, and cascades of instabilities are expected, but their endpoint is an open problem \cite{Niehoff:2015oga}. 
A study of such of nonlinear evolution is \cite{Chesler:2018txn}.

In holography, time periodic states have been studied by applying periodic driving by external sources at the conformal boundary \cite{Li:2013fhw,Natsuume:2013lfa,Auzzi:2013pca,Rangamani:2015sha,Hashimoto:2016ize,Kinoshita:2017uch,Biasi:2017kkn,Ishii:2018ucz,Biasi:2019eap}.
Such time dependent solutions provide a time dependent oscillating condensate called \textit{Floquet condensate} in the dual field theory. 
In the presence of a black hole horizon, the driving can be dissipated; when a horizon is absent, resonances can be excited. 
These driven states survive in the limit of zero source amplitude, resulting in spontaneously excited states which nonlinearly extend the normal modes. 
In this paper, we directly construct cohomogeneity-1 solutions without applying a nontrivial boundary source.

This paper is organized as follows. In Section~\ref{sec:adssoliton}, we study spin-2 normal modes of the AdS soliton. In Section~\ref{sec:nonlinear}, we set up the nonlinear extension of the normal modes. Numerical results for the resonating AdS soliton are presented in Section~\ref{sec:results}. This paper is concluded with a summary and discussion in Section~\ref{sec:conclusion}.

\section{Normal modes of AdS soliton}
\label{sec:adssoliton}

\subsection{AdS soliton}
We consider five dimensional Einstein gravity with a negative cosmological constant,
\begin{equation}
 R_{\mu\nu}-\frac{1}{2}g_{\mu\nu}R = \frac{6}{L^2}g_{\mu\nu}\ ,
\label{Ein}
\end{equation}
where $L$ is the AdS radius.
We will set $L=1$ throughout this paper.

The AdS soliton is a solution to the Einstein equation~(\ref{Ein}) 
with a flat conformal boundary:
\begin{equation}
 ds^2=\frac{1}{z^2}\left[
-dt^2+\frac{dz^2}{F(z)}+\frac{z_0^2}{4}F(z)d\theta^2+dx^2+dy^2
\right]\ ,\quad F(z)=1-\frac{z^4}{z_0^4}\ .
\label{AdSsoliton}
\end{equation}
This metric can also be obtained by the double Wick rotation of the Schwarzschild-AdS$_5$ solution with a flat horizon~\cite{Horowitz:1998ha}.
The AdS boundary is located at $z=0$.
The coordinate $\theta$ is compactified on a circle with a periodicity $\theta \simeq \theta+2\pi$ in order to avoid a conical singularity at the tip of the geometry where $z=z_0$.
This geometry contains 3d Minkowski space denoted by $-dt^2+dx^2+dy^2$ and is invariant under $ISO(2,1)$.
It is also invariant under a $\theta$-translation. 
In summary, the isometry group of the AdS soliton is
\begin{equation}
 ISO(2,1) \times U(1)\ .
\label{isometryAdSsoliton}
\end{equation}
For later convenience, we denote the generator of the rotation in the $(x,y)$-plane by\footnote{
In the polar coordinates $x=\rho \cos\phi$ and $y=\rho\sin\phi$; this is simply written as $\xi=\partial_\phi$.
}
\begin{equation}
 \xi = x\frac{\partial}{\partial y} - y\frac{\partial}{\partial x} \ .
\end{equation}
This is the generator of $U(1)\in ISO(2,1)$.

The compactified direction introduces the Kaluza-Klein mass scale $M_{KK}$.
Near the AdS boundary $z\sim 0$, the metric (\ref{AdSsoliton}) becomes
\begin{equation}
 ds^2\simeq \frac{1}{z^2}\left[-dt^2+dx^2+dy^2+\frac{z_0^2}{4}d\theta^2\right]\ .
\end{equation}
The boundary metric is locally 4d Minkowski, but the $\theta$-direction is compactified.
For the AdS soliton, the Kaluza-Klein mass scale is given by
\begin{equation}
 M_{KK}=\frac{2}{z_0}\ .
 \label{MKK_AdSsoliton}
\end{equation}
In this paper, we will measure dimensionful quantities in units of $M_{KK}$.

\subsection{Decoupled spin-2 perturbation}

We focus on the decoupled spin-2 gravitational perturbation of the AdS soliton that is homogeneous in the $(x,y)$-plane \cite{Constable:1999gb,Brower:1999nj}.
Here, we also include a nontrivial Kaluza-Klein momentum $k$ along the $\theta$-coordinate.
We formulate the perturbation in a way that will naturally leads to the nonlinear construction in the following section.

Let us introduce complex coordinates $w_\pm$ in the $(x,y)$-plane as
\begin{equation}
 dw_\pm = dx \pm i dy\ .
\end{equation}
These satisfy
\begin{equation}
\mathcal{L}_\xi dw_\pm = \pm i dw_\pm \ , 
\label{dwcharge}
\end{equation}
where $\mathcal{L}$ is the Lie derivative. 
This means that $dw_\pm$ have $U(1)$-charges of $\pm 1$.
Then we consider the following gravitational perturbation of the AdS soliton:
\begin{equation}
 h_{\mu\nu}dx^\mu dx^\nu
= \frac{1}{z^2}e^{-i\omega t + ik \theta} \delta \alpha(z) dw_+^2 + \textrm{h.c.}\ .
\label{da_pert}
\end{equation}
From the periodicity of $\theta$, $k\in \bm{Z}$ is required. 
This perturbation has $U(1)$-charges of $+2$ (first term) and $-2$ (second term), and it is homogeneous in the $(x,y)$-plane: 
$\partial_x h_{\mu\nu}=\partial_y h_{\mu\nu}=0$.
Other homogeneous perturbations such as $dt^2, \, dtdw_+$, and $dz dw_-$ cannot have $U(1)$-charges of $\pm 2$.
Thus the perturbation~(\ref{da_pert}) is decoupled from the other perturbations.

The perturbation equation for $\delta \alpha$ is given by
\begin{equation}
 \delta \alpha'' + \frac{(z^{-3}F)'}{z^{-3}F} \delta \alpha'
+\left(\frac{\omega^2}{F} - \frac{4 k^2}{z_0^2 F^2} - \frac{2(zF'-4F+4)}{z^2F} \right) \delta \alpha = 0\ .
\label{perteq}
\end{equation}
Solving the above equation near the tip $z=z_0$ and conformal boundary $z=0$, we find the 
regular and Dirichlet boundary conditions as
\begin{equation}
\delta \alpha\sim (z_0-z)^{k/2} \quad (z\sim z_0)\ ,
\qquad 
 \delta \alpha\sim z^4\quad (z\sim 0)\ .
\end{equation}
We compute the normal mode spectrum for $\omega$ by solving \eqref{perteq} numerically with these boundary conditions.
Results are summarized in Table~\ref{tab:spec}. There, $n$ denotes the radial overtone number given by the number of nodes in the interval $0 \le z \le z_0$ and corresponds to the excitations of the dual spin-2 glueballs \cite{Constable:1999gb,Brower:1999nj}.
The numerical values for $k=0$ reproduce the spectrum obtained in \cite{Csaki:1998qr,deMelloKoch:1998vqw,Zyskin:1998tg} for $0^{++}$ glueballs, which degenerate with the $2^{++}$ glueball spectrum \cite{Constable:1999gb,Brower:1999nj}.

\begin{table}[t]
\begin{center}
\caption{The spectrum of the gravitational perturbation $\omega/M_{KK}$.}
\vspace{1.5mm}
  \begin{tabular}{|c|c|c|c|c|} \hline
       & $n=0$  & $n=1$   & $n=2$ & $n=3$ \\ \hline 
   $k=0$ & 1.7020 &2.9380 &4.1526 &5.3598 \\
   $k=1$ & 2.4224 &3.6144 &4.8082 &6.0034 \\
   $k=2$ & 3.2544 &4.3848 &5.5424 &6.7139 \\ 
   $k=3$ & 4.1412 &5.2130 &6.3309 &7.4740 \\ 
   $k=4$ & 5.0587 &6.0788 &7.1580 &8.2717 \\ \hline 
  \end{tabular}
\label{tab:spec}
\end{center}
\end{table}

\section{Nonlinear resonation of AdS soliton}
\label{sec:nonlinear}

\subsection{Symmetry preserved by the perturbation}

The main goal of this paper is to construct nonlinear solutions that extend from the normal modes obtained in the previous section. 
In this section, we identify the metric ansatz for the nonlinear solutions and set up its computation.
A perturbative construction has been given in \cite{Hartnett:2012np}, but here we will construct complete nonlinear solutions. In particular, we find that they are obtained in a cohomogeneity-1 metric ansatz.

To find the metric ansatz for the nonlinear solutions, we examine the symmetries preserved by the perturbation~(\ref{da_pert}).
Let us define the following 1-forms $e_\pm$ by
\begin{equation}
 e_\pm = e^{\pm (-i\omega t + ik\theta)/2} dw_\pm\ .
\end{equation}
In terms of them, the perturbation~(\ref{da_pert}) is simply written as
\begin{equation}
 h_{\mu\nu}dx^\mu dx^\nu= \frac{1}{z^2}\delta \alpha (z) (e_+^2+e_-^2)\ .
\label{da_pert2}
\end{equation}
Hence it is convenient to look at the symmetries acting on $e_\pm$.
The 1-forms $e_\pm$ are transformed by $\xi$, $\partial_t$ and $\partial_x$ as
\begin{equation}
\mathcal{L}_\xi e_\pm = \pm i e_\pm\ ,\quad
\mathcal{L}_{\partial_t} e_\pm = \mp \frac{i\omega}{2} e_\pm\ ,\quad
\mathcal{L}_{\partial_\theta} e_\pm = \pm \frac{ik}{2} e_\pm\ .
\end{equation}
The $t$ and $\theta$-translations are broken if $\omega \neq 0$ and $k \neq 0$, respectively, but their linear combinations with $\xi$ are preserved, defining vector fields $K_1$ and $K_2$ by
\begin{equation}
 K_1 = \partial_t +\frac{\omega}{2}\xi\ ,\qquad
 K_2= \partial_\theta -\frac{k}{2}\xi\ ,
\end{equation}
implying that
\begin{equation}
\mathcal{L}_{K_1} e_\pm = 0\ ,\qquad 
\mathcal{L}_{K_2} e_\pm = 0\ .
\end{equation}
Thus, the perturbation~(\ref{da_pert2}) preserves the symmetries generated by $K_1$ and $K_2$.
That is, $\xi$, $\partial_t$ and $\partial_x$ are not Killing vectors independently, 
but their appropriate linear combinations are.
In $K_1$, the $t$-translation is combined with the rotation generated by $U(1)_\xi$, and therefore $K_1$ can be regarded as a helical Killing vector.
The 1-forms $e_\pm$ are also invariant under translations in the $(x,y)$-plane:
$\mathcal{L}_{\partial_x} e_{\pm} = \mathcal{L}_{\partial_y} e_{\pm}=0$.
The upshot is that the perturbation~(\ref{da_pert2}) is invariant under $\{K_1, K_2, \partial_x, \partial_y\}$.

What is the isometry group generated by $\{K_1, K_2, \partial_x, \partial_y\}$?
To identify that, it is more appropriate to consider $L_1 \equiv 2K_1/\omega = \xi + (2/\omega)\partial_t$ and $L_2 \equiv k K_1 + \omega K_2=k\partial_t + \omega \partial_\theta$ so that $\xi$ and $\partial_\theta$ are separated. Clearly, $\mathcal{L}_{L_1} e_\pm = 0 = \mathcal{L}_{L_2} e_\pm$.
Recall that $\{\xi,\partial_x, \partial_y\}$ form the algebra of $ISO(2)$, although $\xi$ is no longer a Killing vector.
Then, because $[\partial_t, \partial_x]=[\partial_t, \partial_y]=0$, we can say that
$\{L_1,\partial_x, \partial_y\}$ are the generators of $ISO(2)$.
The remaining one $L_2$ commutes with the other vectors $\{L_1,\partial_x, \partial_y\}$ and 
is the generator of $U(1)$ for $k=0$ and $R$ for $k\neq 0$.
Thus, the isometry group generated by $\{K_1, K_2, \partial_x, \partial_y\}$ is given by\footnote{If $k=0$, $L_2$ does not depend on $K_1$, while if $k \neq 0$ both $L_1$ and $L_2$ depends on $K_1$. Therefore, the helical isometry given by $K_1$ is contained in $ISO(2)$ for $k=0$ and distributed in $ISO(2) \times U(1)$ for $k \neq 0$.}
\begin{equation}
\begin{split}
 &ISO(2)\times U(1)\qquad (k=0)\ , \\
 &ISO(2)\times R\qquad (k\neq0)\ .
\end{split}
\label{isometryG}
\end{equation}
The original isometry group of the AdS soliton~(\ref{isometryAdSsoliton}) 
is broken into these smaller groups by the perturbation~(\ref{da_pert2}).

The perturbation~(\ref{da_pert2}) is also invariant under discrete transformations $P_1$ and $P_2$ given by
\begin{equation}
\begin{split}
 &P_1(t,\theta,x,y) = (-t,-\theta,-x,y)\ ,\\
 &P_2(t,\theta,x,y) = (-t,-\theta,x,-y)\ .
\end{split}
\label{parity_xy}
\end{equation}
Under these, 1-forms $(dt,d\theta,e_+,e_-)$ are transformed as
\begin{equation}
\begin{split}
 &P_1(dt,d\theta,e_+,e_-) = (-dt,-d\theta,-e_-,-e_+)\ ,\\
 &P_2(dt,d\theta,e_+,e_-) = (-dt,-d\theta,e_-,e_+)\ .
\end{split}
\label{parity_e}
\end{equation}

\subsection{Metric for resonating AdS soliton}

We will now construct the metric ansatz that has Killing vectors $\{K_1, K_2, \partial_x, \partial_y\}$.
To this end, it is convenient to  introduce real 1-forms $e_1$ and $e_2$ as
\begin{equation}
 e_\pm = e_1 \pm i e_2\ .
\end{equation}
They can be related to the original orthogonal coordinates, $(x,y)$, as
\begin{equation}
 \begin{pmatrix}
 e_1 \\
 e_2
 \end{pmatrix}
=  
\begin{pmatrix}
\cos\Theta &  -\sin\Theta\\
\sin\Theta  & \cos\Theta
\end{pmatrix}
\begin{pmatrix}
 dx \\
 dy
\end{pmatrix}
\ ,\qquad
\Theta = \frac{1}{2}(-\omega t + k \theta)\ .
\end{equation}
These are also invariant under $\{K_1, K_2, \partial_x, \partial_y\}$.
Then, we can write a general metric equipped with them as
\begin{equation}
ds^2 = g_{ab}(z)E^a E^b\ ,
\label{generalmetric}
\end{equation}
where $E^a=(dt,dz,d\theta,e_1,e_2)$. 
The metric depends only on $z$ but still has 15 components.
This can be further simplified by imposing the parity \eqref{parity_e}, under which $(dt,d\theta,e_1,e_2)$ are transformed as
\begin{equation}
\begin{split}
 &P_1(dt,d\theta,e_1,e_2) = (-dt,-d\theta,-e_1,e_2)\ ,\\
 &P_2(dt,d\theta,e_1,e_2) = (-dt,-d\theta,e_1,-e_2)\ .
\end{split}
\end{equation}
Therefore, in Eq.~(\ref{generalmetric}), only the terms with $dt^2, \, dtd\theta, \, d\theta^2, \, dz^2, \, e_1^2,$ and $e_2^2$ are allowed if the parity invariance is imposed.
This leads us to write the cohomogeneity-1 metric ansatz as
\begin{multline}
 ds^2=\frac{1}{z^2}\bigg[
-f(z)dt^2+\frac{dz^2}{F(z)g(z)}+\alpha(z) e_1^2 + \frac{1}{\alpha(z)} e_2^2\\
+\frac{z_0^2}{4}F(z)\beta(z)\{d\theta-h(z)dt\}^2
\bigg]\ ,
\label{coh-1}
\end{multline}
where $F(z)$ was defined in Eq.~(\ref{AdSsoliton}). The product of the coefficients of $e_1^2$ and $e_2^2$ are fixed by redefinition of the $z$-coordinate. Note that a similar metric has also appeared as an effective metric on the probe D7-brane when a rotating electric field is applied in the D3/D7 system~\cite{Hashimoto:2016ize,Kinoshita:2017uch}.
To avoid a conical singularity at the tip, we require
\begin{equation}
 g(z)\beta(z)|_{z=z_0}=1 \ .
\label{regcond}
\end{equation}
The AdS soliton (\ref{AdSsoliton}) is reproduced when $f=g=\alpha=\beta=1$ and $h=0$.

Substituting the ansatz (\ref{coh-1}) into the Einstein field equation $G_{\mu\nu}=6g_{\mu\nu}$, we obtain the complete set of the equations of motions as
\begin{align}
 f'=&\frac{1}{4  z_0^2 z F g \alpha^2  \{z(F \beta)'-6F\beta)\}}\big[
z_0^2 F \{
z^2 F g \beta(-z_0^2 F h'{}^2 \alpha^2  \beta 
+ 4 \alpha'{}^2 f) \notag\\
&+ 24 zfg \alpha^2 (F \beta)'
- 96 f \alpha^2 \beta(Fg-1)
\}\notag
\\
&+z^2 (\alpha^2 - 1)^2  \{ z_0^2 F \beta (\omega-kh)^2  - 4 k^2 f\}
\big]
\ , \label{feq}\\
g'=&\frac{1}{4 z0^2z  F f \alpha^2 \beta(z (F \beta)'-6F\beta)}
\big[-z_0^2 \{
z^2g\alpha^2 \beta^2 (-z_0^2  F^3  h'{}^2 \beta + 8 F'{}^2 f) \notag\\
&+ 4 zF'Ffg \alpha^2\beta(3  z \beta' - 14  \beta ) 
+ 4 z^2 F^2 f g (\alpha^2 \beta'{}^2   + \alpha'{}^2 \beta^2) \notag\\
&+32zf\alpha^2\beta (-F^2 g \beta'+(F \beta)')+ 96 Ff\alpha^2\beta^2 (gF-1)
\} \notag\\
&+z^2 \beta  (\alpha^2 - 1)^2 (-z_0^2 F \beta (\omega-kh)^2  + 4 k^2 f)
\big]
\ ,\label{geq}\\
h''=&\frac{h'}{2}\left(
\frac{f'}{f}-\frac{g'}{g}-\frac{3\beta'}{\beta}-\frac{4F'}{F}+\frac{6}{z}
\right)-\frac{k(\omega-kh)}{z_0^2 F^2 g \beta}\left(\alpha-\frac{1}{\alpha}\right)^2
\ ,\label{heq}\\
\alpha''=&\alpha'\left(\frac{\alpha'}{\alpha}+\frac{4}{zFg}-\frac{1}{z}\right)
-\frac{\alpha (z_0^2 F \beta (\omega-kh)^2  - 4 k^2 f)}{4 z_0^2  F^2 f g \beta}\left(\alpha^2-\frac{1}{\alpha^2}\right)
\ ,\label{aeq}\\
\beta''=&-\frac{(Fg-4)(F\beta)'}{z F^2 g} \notag\\
&+\beta\left(\frac{\beta'{}^2}{\beta^2} - \frac{F''}{F} + \frac{F'{}^2}{F^2} -\frac{z_0^2 Fh'{}^2 \beta}{4 f} \right)
-\frac{k^2}{z_0^2 F^2 g}\left(\alpha-\frac{1}{\alpha}\right)^2
\ ,\label{beq}
\end{align}
where ${}' \equiv d/dz$.
We will solve them numerically.
Note that $h$ decouples from the rest when $k=0$.

\subsection{Asymptotic form at the tip}
The asymptotic solution to (\ref{feq}-\ref{beq}) near the tip $z=z_0$ with the regularity condition~(\ref{regcond}) takes the form
\begin{equation}
\begin{split}
 &\alpha=1+r^k(\alpha_0 + \alpha_1 r + \alpha_2 r^2+\cdots)\ ,\\
 &g=1+r^{2k}(g_0 + g_2 r^2 + g_4 r^4+\cdots)\ ,\\
 &\beta=1+r^{2k}(\beta_0 + \beta_2 r^2 + \beta_4 r^4+\cdots)\ ,\\
 &f=F_0+r^{2k+2}(f_0 + f_2 r^2 + f_4 r^4+\cdots)\ ,\\
 &h=H_0+r^{2k}(h_0 + h_2 r^2 + h_4 r^4+\cdots)\ ,
\end{split}
\label{neartip}
\end{equation}
where $r^2\equiv (z-z_0)/z_0$. Once we specify five parameters $(\alpha_0,F_0,H_0)$ and $(\omega,z_0)$, 
other expansion coefficients $(\alpha_i,\beta_i,f_i,g_i, h_i)$ are determined by the equations of motion.
For example, the leading order coefficients are
\begin{equation}
\begin{split}
 &f_0=\frac{z_0^2(\omega-kH_0)^2}{4(k+1)^2}\left(1-\frac{\alpha_0(3\alpha_0+4)}{4(\alpha_0 + 1)^2)}\delta_{k,0}\right)\alpha_0^2\ ,\quad
 g_0=\frac{k^2}{4(k+1)^2}\alpha_0^2\ ,\\
 &h_0=-\frac{\omega-kH_0}{4(k+1)}(1-\delta_{k,0})\alpha_0^2\ ,\quad
 \beta_0=-\frac{k(k+2)}{4(k+1)^2}\alpha_0^2\ .
\end{split}
\end{equation}
Because the field redefinition $\alpha\to 1/\alpha$ just exchanges the roles of $e_1$ and $e_2$, 
we can assume $\alpha_0\geq 0$ without loss of generality.

\subsection{Physical quantities}

At the AdS boundary, we impose the asymptotically locally AdS condition as
\begin{equation}
 f\to 1\ ,\qquad \alpha \to 1\ , \qquad h\to 0\ ,\qquad (z\to 0)\ ,
\label{ato1}
\end{equation}
while $g\to 1 \ (z\to 0)$ is automatically satisfied because of the equations of motion.
Solving Eqs.~(\ref{feq}-\ref{beq}) near $z=0$ with the above boundary condition gives the asymptotic solution as
\begin{equation}
\begin{split}
f&=1 + c_f \left(\frac{z}{z_0}\right)^4+\cdots\ ,\quad 
g=1 + (c_f + c_\beta) \left(\frac{z}{z_0}\right)^4+\cdots\ ,\\
h&= \frac{c_h}{z_0} \left(\frac{z}{z_0}\right)^4+\cdots\ ,\quad
\alpha=1+c_\alpha \left(\frac{z}{z_0}\right)^4+\cdots\ ,\\
\beta&=\beta_\infty \left[1+ c_\beta \left(\frac{z}{z_0}\right)^4+\cdots\right]\ ,
\end{split}
\label{fghab_asym}
\end{equation}
where $\beta_\infty$, $c_f$, $c_h$, $c_\alpha$ and $c_\beta$ are unspecified in the asymptotic analysis.
These will be determined when the equations of motion are solved in the bulk.
The metric of the conformal boundary is
\begin{equation}
 ds_{\textrm{bdry}}^2=-dt^2+dx^2+dy^2+\frac{d\theta^2}{M_{KK}{}^2} \ ,
\end{equation}
where the Kaluza-Klein mass scale for the metric (\ref{coh-1}) is given by\footnote{For the undeformed AdS soliton, $\beta_\infty=1$ and (\ref{MKK_AdSsoliton}) is reproduced.}
\begin{equation}
M_{KK}=\frac{2}{z_0\sqrt{\beta_\infty}}\ .
\end{equation}
Let us introduce the canonically normalized 
coordinate $\chi=\theta/M_{KK}$, which has periodicity $\chi\simeq \chi+2\pi/M_{KK}$.
Then the boundary metric is rewritten as
\begin{equation}
 ds_{\textrm{bdry}}^2=-dt^2+dx^2+dy^2+d\chi^2\ .
 \label{dsbdry_chi}
\end{equation}

Thermodynamical quantities are obtained from the boundary energy momentum tensor, which is given by \cite{Ashtekar:1999jx,Kinoshita:2008dq}\footnote{Also see the counterterm method \cite{Balasubramanian:1999re,deHaro:2000vlm,Bianchi:2001kw}.}
\begin{equation}
 8\pi G_5 T_{ij}=-\frac{1}{2z^2}C_{i\rho j \sigma}n^\rho n^\sigma \bigg|_{z=0}\ ,
\end{equation}
where $i,j=t,x,y,\chi$, $G_5$ is the five dimensional Newton constant, 
$C_{\mu\nu\rho\sigma}$ is the Weyl tensor of the bulk spacetime, and 
$n^\mu$ is the unit normal to a bulk constant-$z$ surface. 
Using the asymptotic solution~(\ref{fghab_asym}), we obtain
\begin{equation}
T_{ij}dx^i dx^j= \frac{M_{KK}^4}{256 \pi G_5}\big[
\mathcal{E}\, dt^2
-2\pi_\chi\, dt d\chi 
-\mathcal{T}\, d\chi^2
+\mathcal{P}\,(e_1^2+e_2^2)
+\sigma\, (e_1^2-e_2^2)
\big]\ ,
\label{Tij}
\end{equation}
where we defined 
\begin{equation}
\begin{split}
 &\mathcal{E}=-\beta_\infty^2 (1+3c_f-c_\beta)\ ,\quad
 \pi_\chi=2\beta_\infty^{5/2} c_h\ ,\quad
 \mathcal{T}=\beta_\infty^2(3+c_f - 3c_\beta)\ ,\\
 &\mathcal{P}=\beta_\infty^2(1-c_f -c_\beta)\ ,\quad
 \sigma = 4 \beta_\infty^2 c_\alpha\ .
\end{split}
\label{Thq}
\end{equation}
These quantities are interpreted as\footnote{We define these quantities in the viewpoint of the (3+1)-dimensional boundary theory.
By dimensional reduction along the $\chi$-direction,
effective (2+1)-dimensional quantities can be obtained as $\mathcal{E}_{(2+1)d}=(2\pi/M_{KK}) \mathcal{E}$, etc.}
\begin{itemize}
 \item $\mathcal{E}$: Energy density 
 \item $\pi_\chi$: Momentum density along the compact direction
 \item $\mathcal{T}$: Tension along the compact direction (i.e.~Casimir force)
 \item $\mathcal{P}$: (Time average of) pressure
 \item $\sigma$: (The maximal value of the) shear stress
\end{itemize}
up to the normalization factor 
\begin{equation}
 \frac{M_{KK}^4}{256 \pi G_5} = \frac{M_{KK}^4 N_c^2}{128\pi^2}\ ,
\end{equation}
where $N_c$ is the number of colors in the dual gauge theory.
For the AdS soliton, we obtain $(\mathcal{E},\pi_\chi,\mathcal{T},\mathcal{P},\sigma)=(-1,0,3,1,0).$
The energy density of the AdS soliton is negative~\cite{Horowitz:1998ha}.

In the energy momentum tensor~(\ref{Tij}), only the last term is time dependent. 
(Note that $e_1^2+e_2^2=dx^2+dy^2$ is time independent.)
In the original $(x,y)$-coordinates, the time dependent term is
\begin{equation}
 e_1^2-e_2^2 = \cos(k\theta-\omega t)(dx^2-dy^2)+2\sin(k\theta-\omega t)dxdy\ .
\end{equation}
The $xy$-component of the energy momentum tensor takes the form
\begin{equation}
 T_{xy}\propto \sigma \sin(k\theta-\omega t) \ .
 \label{T_xy_timedepen}
\end{equation}
This is called the shear stress (i.e.~the flux of the $x$-component of the momentum measured on the $(y,\chi)$-plane).
The parameter $\sigma$ gives the maximum value of the shear stress.
Meanwhile, the other part of $\sigma(e_1^2-e_2^2)$ contributes to the $xx$ and $yy$-components of the energy momentum tensor, but the time average of this oscillating contribution vanishes. Therefore, $\mathcal{P}$ is regarded as time averaged pressure.
The energy momentum tensor is time periodic without dissipation. 
This indicates that the resonating AdS soliton is dual to a coherently excited state in the dual field theory.

\subsection{Technical details}

Some tricks are available for solving the equations of motion numerically.
Practically, we use new variables 
$(\tilde{f}, \tilde{g}, \tilde{h}, \tilde{\alpha}, \tilde{\beta})$ introduced as
\begin{equation}
\begin{split}
&f=F_0+r^{2k+2}\tilde{f}(r)\ ,\quad
g=1+r^{2k}\tilde{g}(r)\ ,\quad
h=H_0+r^{2k}\tilde{h}(r)\ ,\\
&\alpha=1+r^k\tilde{\alpha}(r)\ ,\quad
\beta=1+r^{2k}\tilde{\beta}(r)\ .
\end{split}
\end{equation}
where $r^2\equiv (z-z_0)/z_0$. 
We integrate the equations of motion for $(\tilde{f}, \tilde{g}, \tilde{h}, \tilde{\alpha}, \tilde{\beta})$ from $r=0$ ($z=z_0$) to $r=1$ ($z=0$) numerically. 
The boundary condition at $r=0$ is given in Eq.~(\ref{neartip}).
There are five parameters, $(\alpha_0,F_0,H_0,\omega,z_0)$, which we need to specify in the integration prima facie.
We can set $H_0=0$, $F_0=1$ and $z_0=1$ without loss of generality.
Then, we choose a value of $\alpha_0\neq 0$ and determine the frequency $\omega$ 
by the shooting method so that $\alpha\to 1$ is satisfied at infinity.
The solution obtained in this way, however, does not satisfy the other two conditions in Eq.~(\ref{ato1}), i.e.~$f(z=0)=f_\infty\neq 1$ and $h(z=0)=h_\infty\neq 0$.
The solution satisfying Eq.~(\ref{ato1}) can be obtained by using scaling symmetries.
By applying coordinate transformations $t^\textrm{new}=\sqrt{f_\infty} \, t$ and $\theta^\textrm{new}=\theta-h_\infty t^\textrm{new}$, the metric components $f$ and $h$ become
\begin{equation}
f^\textrm{new}(z)=\frac{f(z)}{f_\infty}\ ,\quad
h^\textrm{new}(z)=\frac{1}{\sqrt{f_\infty}}(h(z)-h_\infty)\ .
\end{equation}
These satisfy the desired boundary condition (\ref{ato1}). By this procedure, the frequency $\omega$ is also changed to
\begin{equation}
 \omega^\textrm{new}=\frac{1}{\sqrt{f_\infty}}(\omega-k h_\infty)\ .
\end{equation}
From the canonically normalized solution, we read off the constants 
$\beta_\infty,c_f,c_h,c_\alpha,c_\beta$ as Eq.~(\ref{fghab_asym}) and then obtain thermodynamical quantities from Eq.~(\ref{Thq}).

\section{Results}
\label{sec:results}

\subsection{Physical quantities}

We construct the resonating AdS soliton by increasing the deformation parameter $\alpha_0$, starting from $\alpha_0=0$.
For each frequency $\omega/M_{KK}$ in Table~\ref{tab:spec}, a different family of solutions can be obtained. Here, we will focus on the results for wave numbers $k=0,1,2$ of the fundamental tone $n=0$. We find that, for $ k>0$, deformed solutions cease to exist at finite $\alpha_0$, and if this occurs we terminate computation there. For $k=0$, deformation continues to $\alpha_0 \to \infty$.

\begin{figure}[t]
  \centering
  \subfigure
 {\includegraphics[scale=0.45]{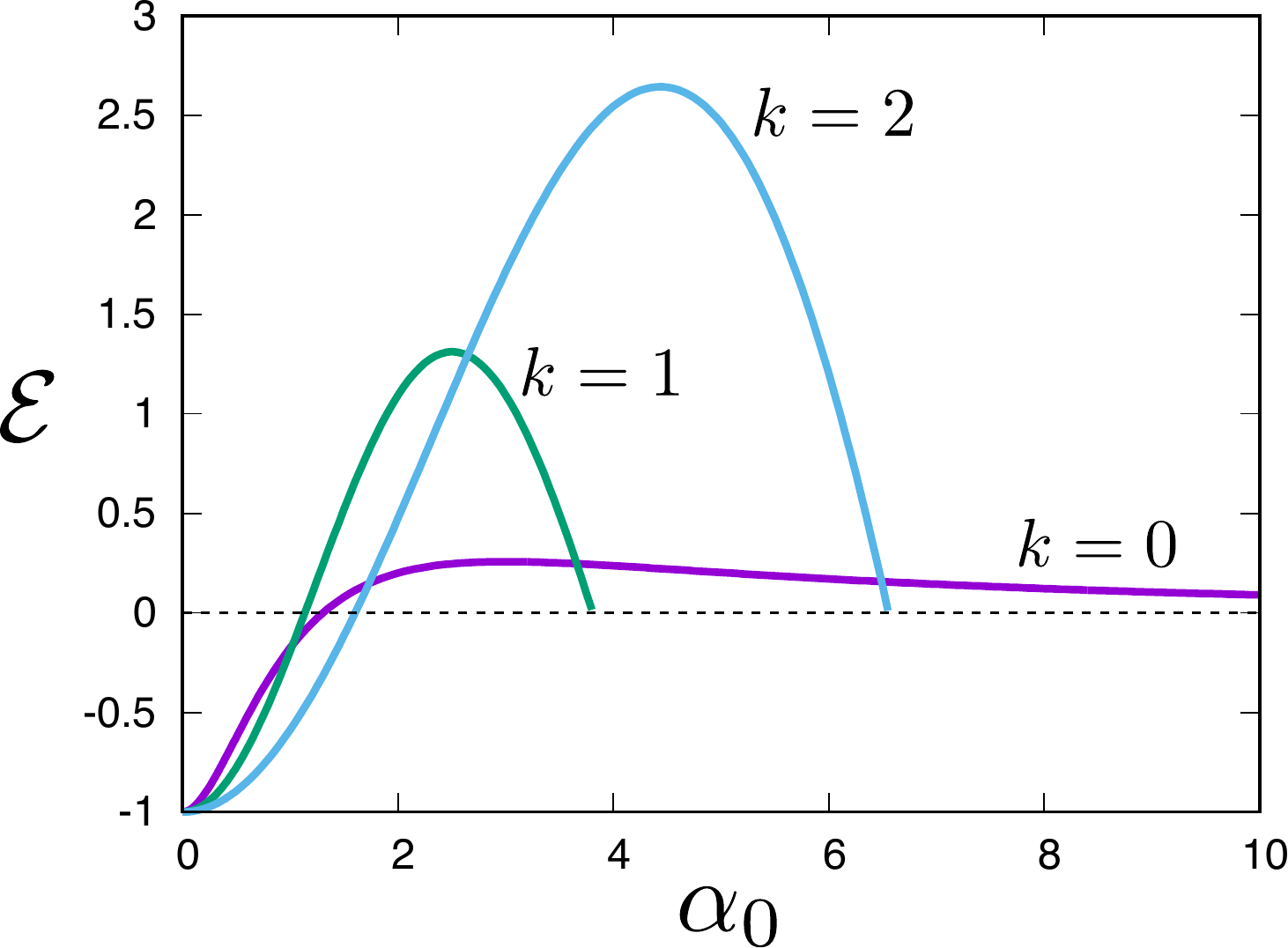}\label{E_vs_a0_k012_n0}
  }
  \subfigure
 {\includegraphics[scale=0.45]{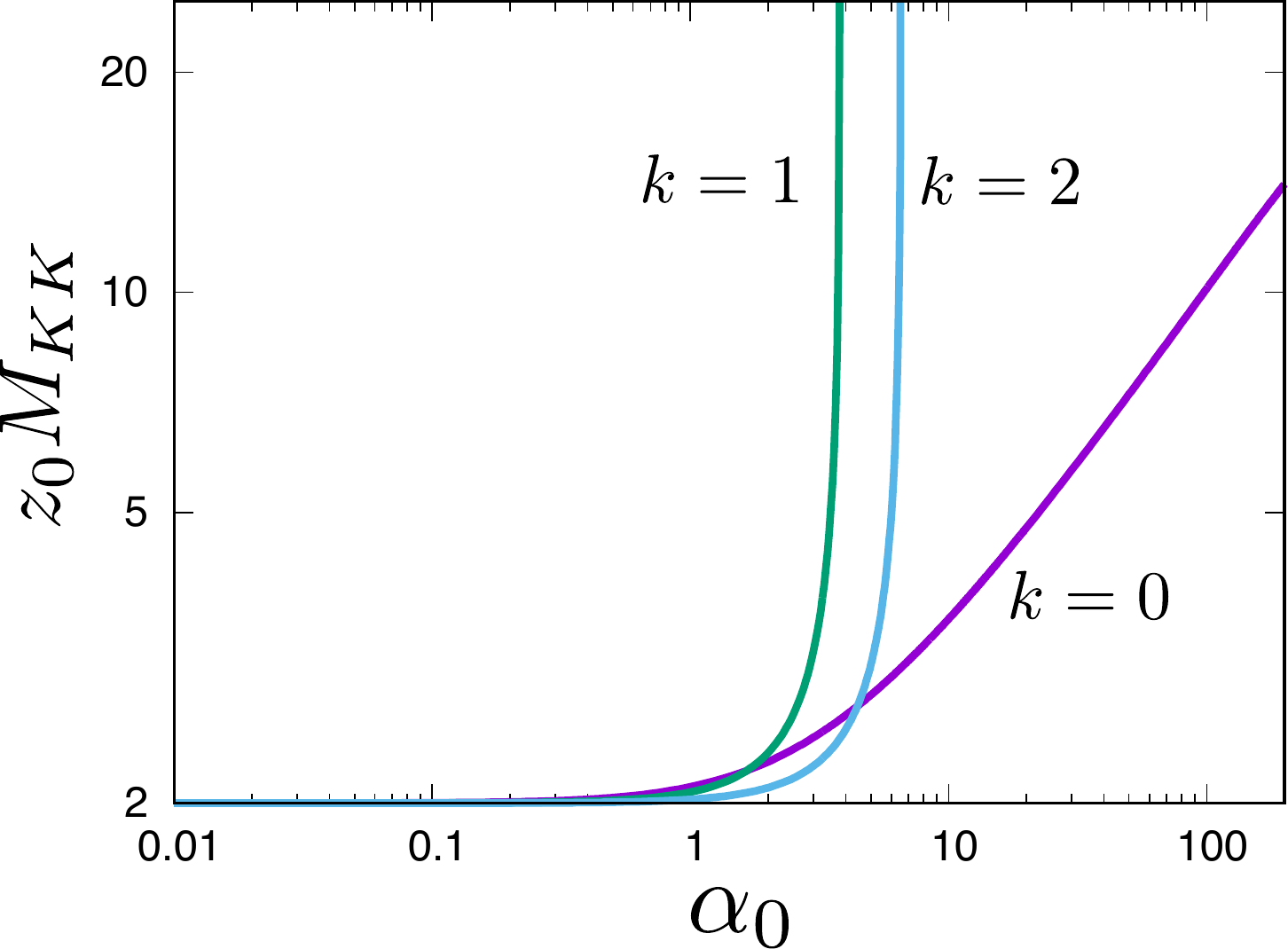}\label{z0_vs_a0_k012_n0}
  }
 \caption{%
(Left) Energy density as a function of $\alpha_0$. 
(Right) Location of the tip of the AdS soliton $z_0$ as a function of $\alpha_0$. 
}
\label{E_and_z0_vs_a0_k0}
\end{figure}

\begin{figure}
  \centering
  \subfigure[Frequency]
 {\includegraphics[scale=0.32]{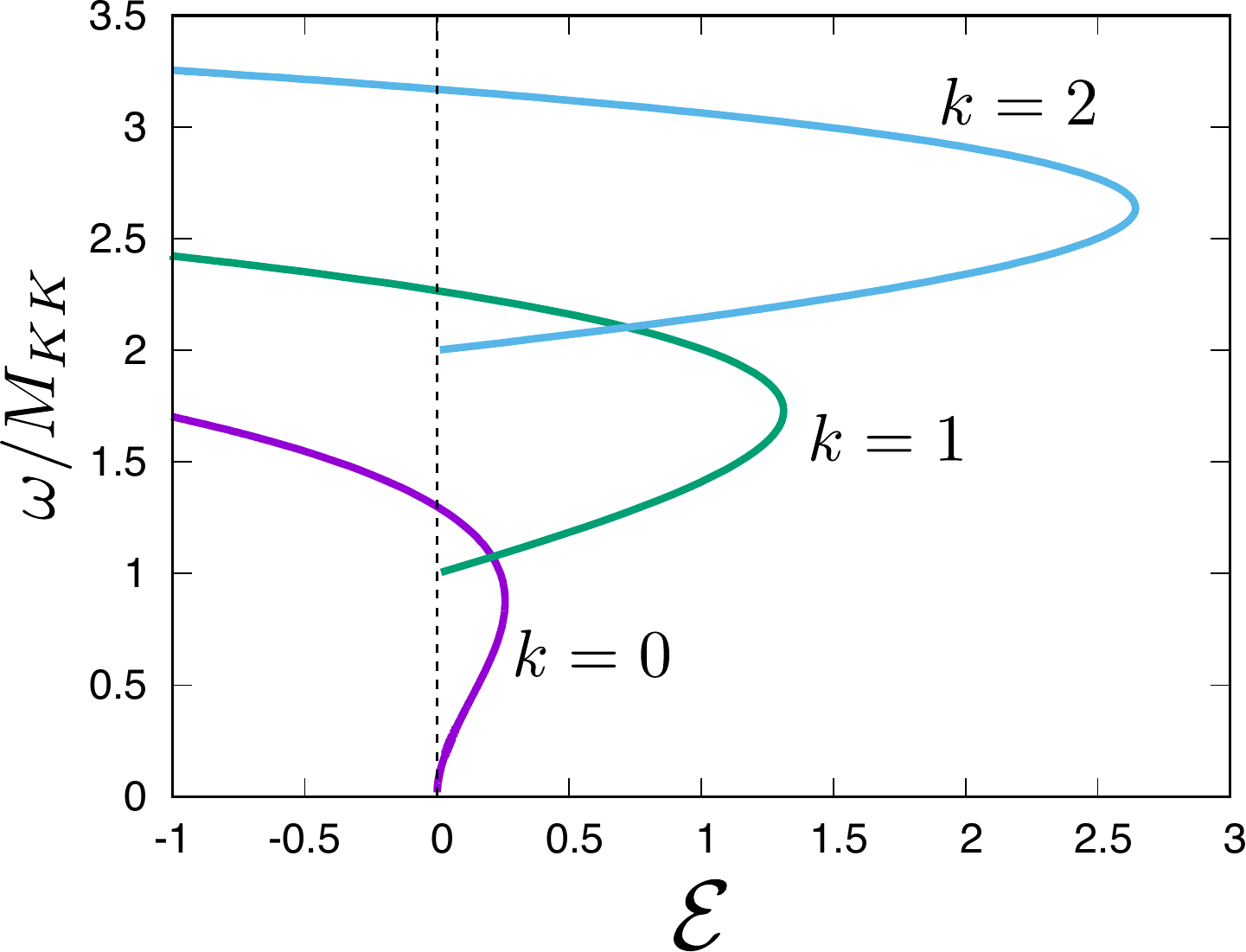}
  }
  \subfigure[Pressure]
 {\includegraphics[scale=0.32]{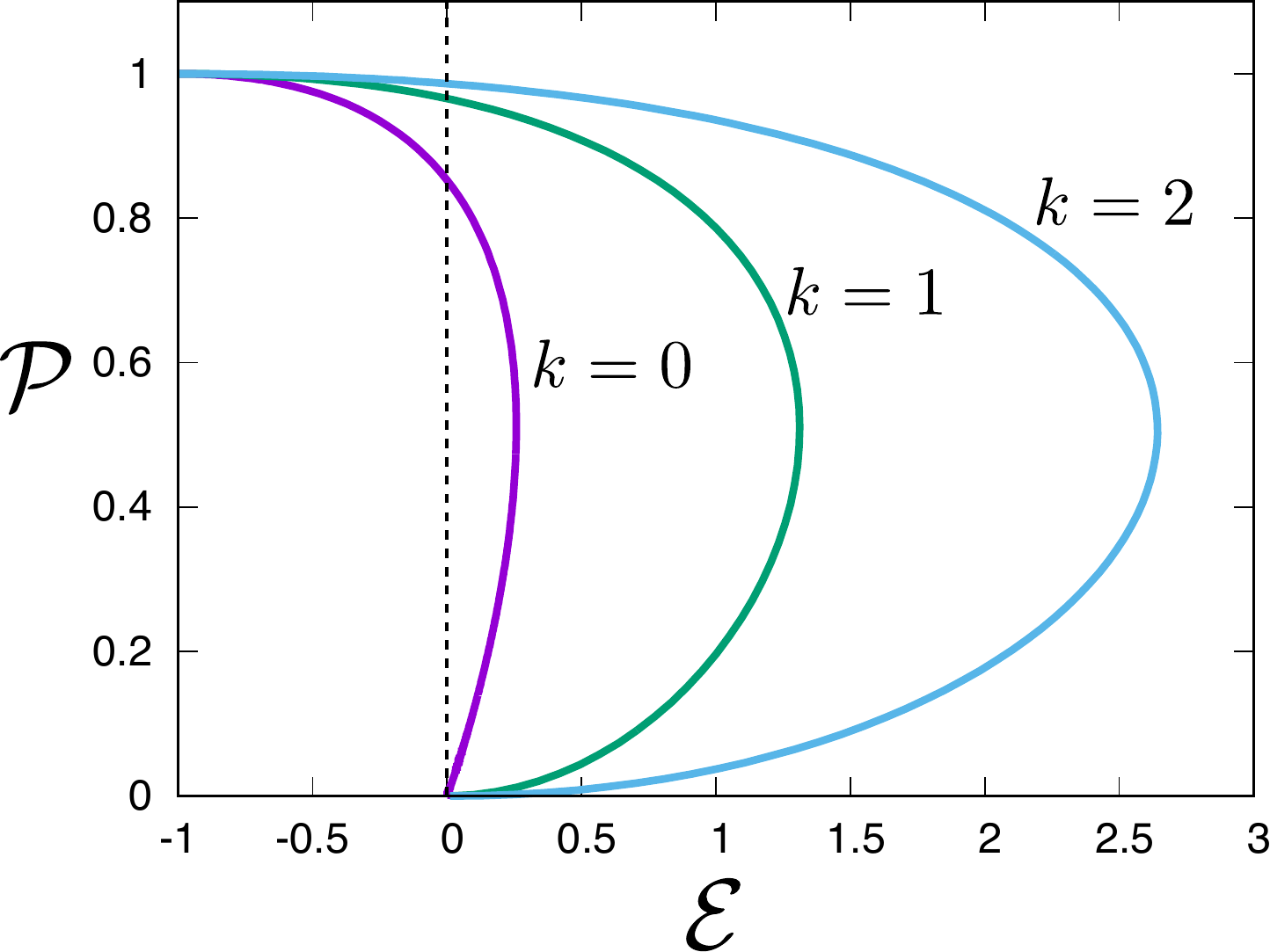}
  }
  \subfigure[Tension]
 {\includegraphics[scale=0.32]{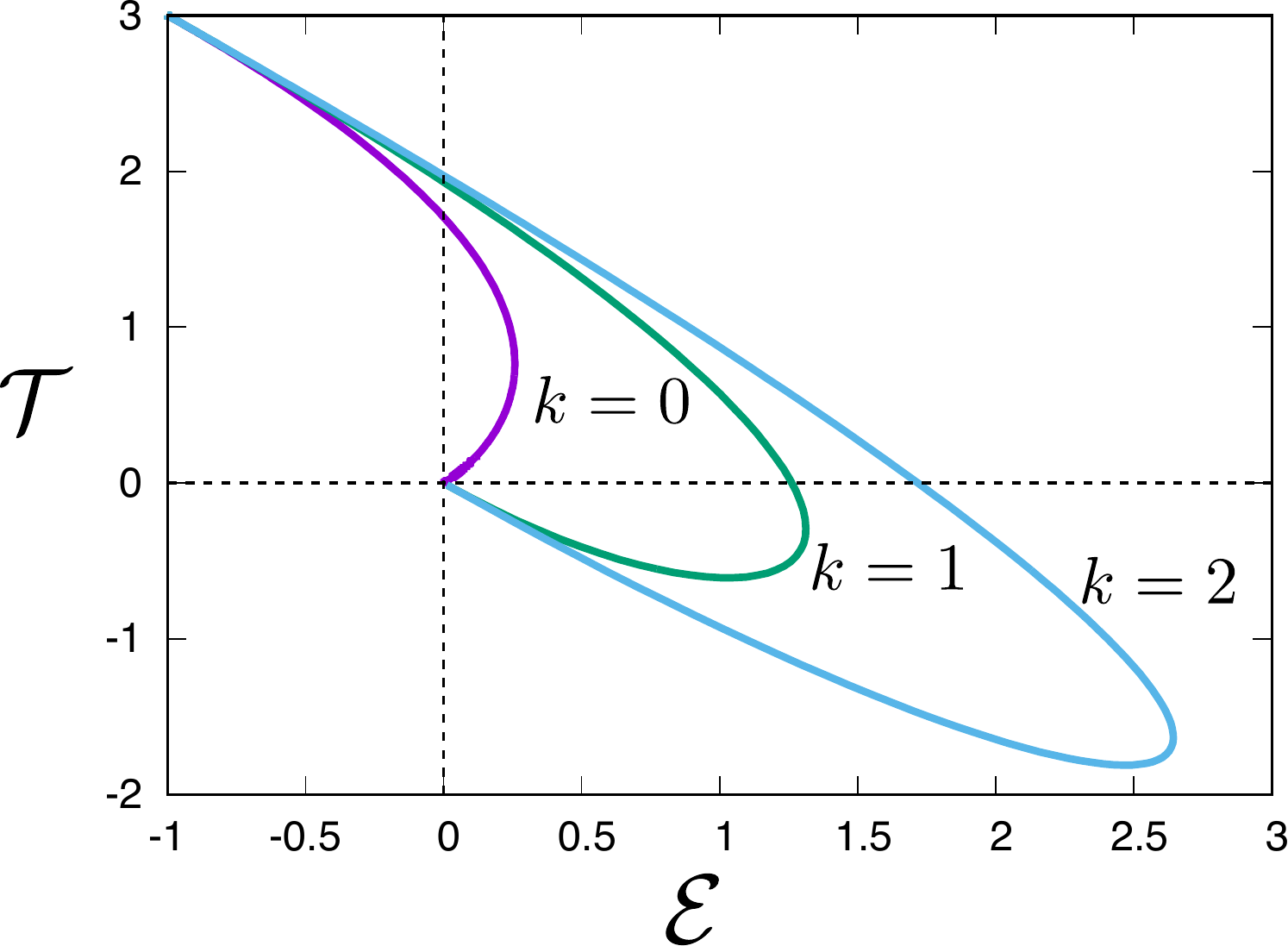}
  }
  \subfigure[Shear stress]
 {\includegraphics[scale=0.32]{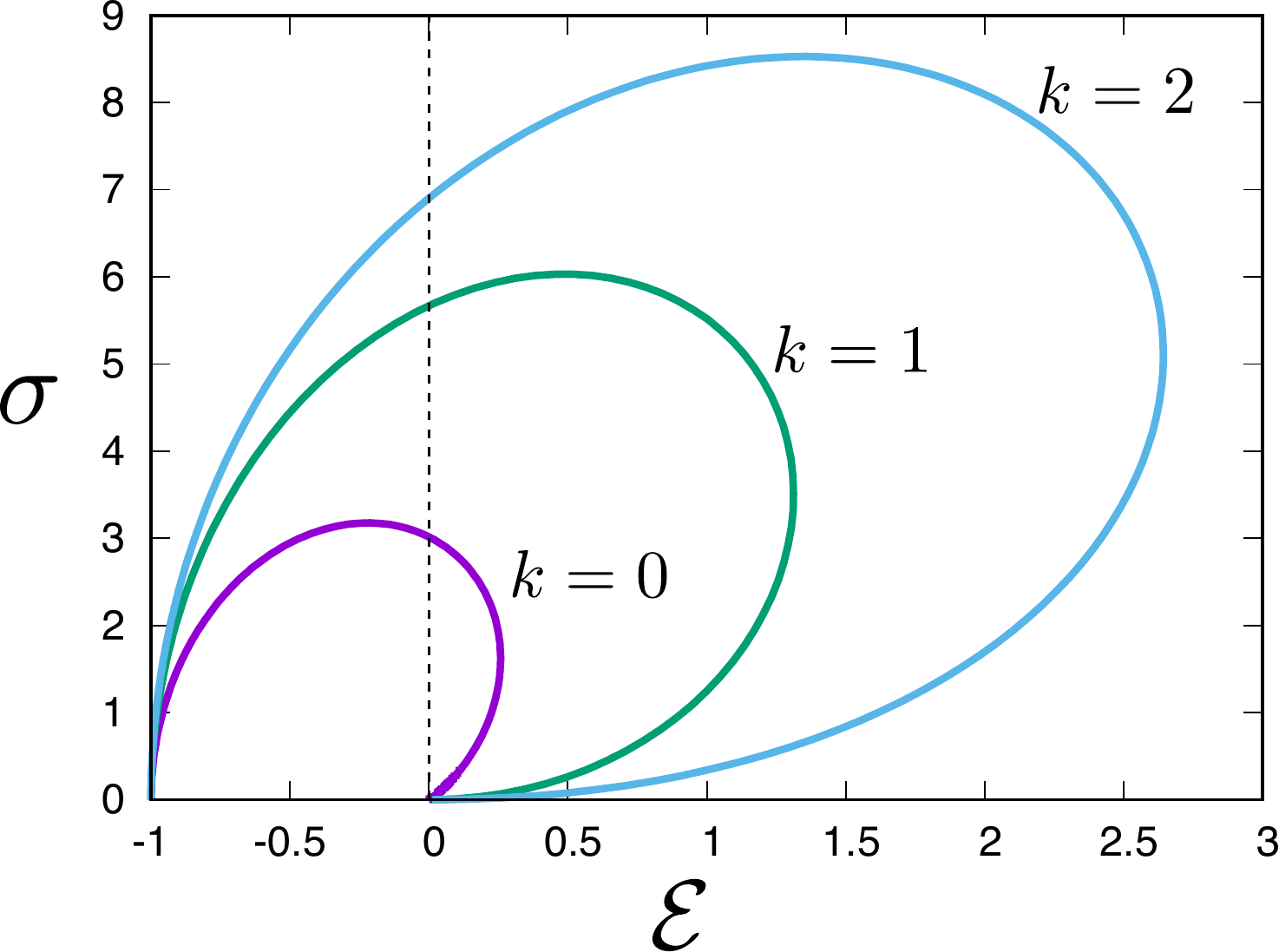}
  }
\subfigure[Momentum]
 {\includegraphics[scale=0.32]{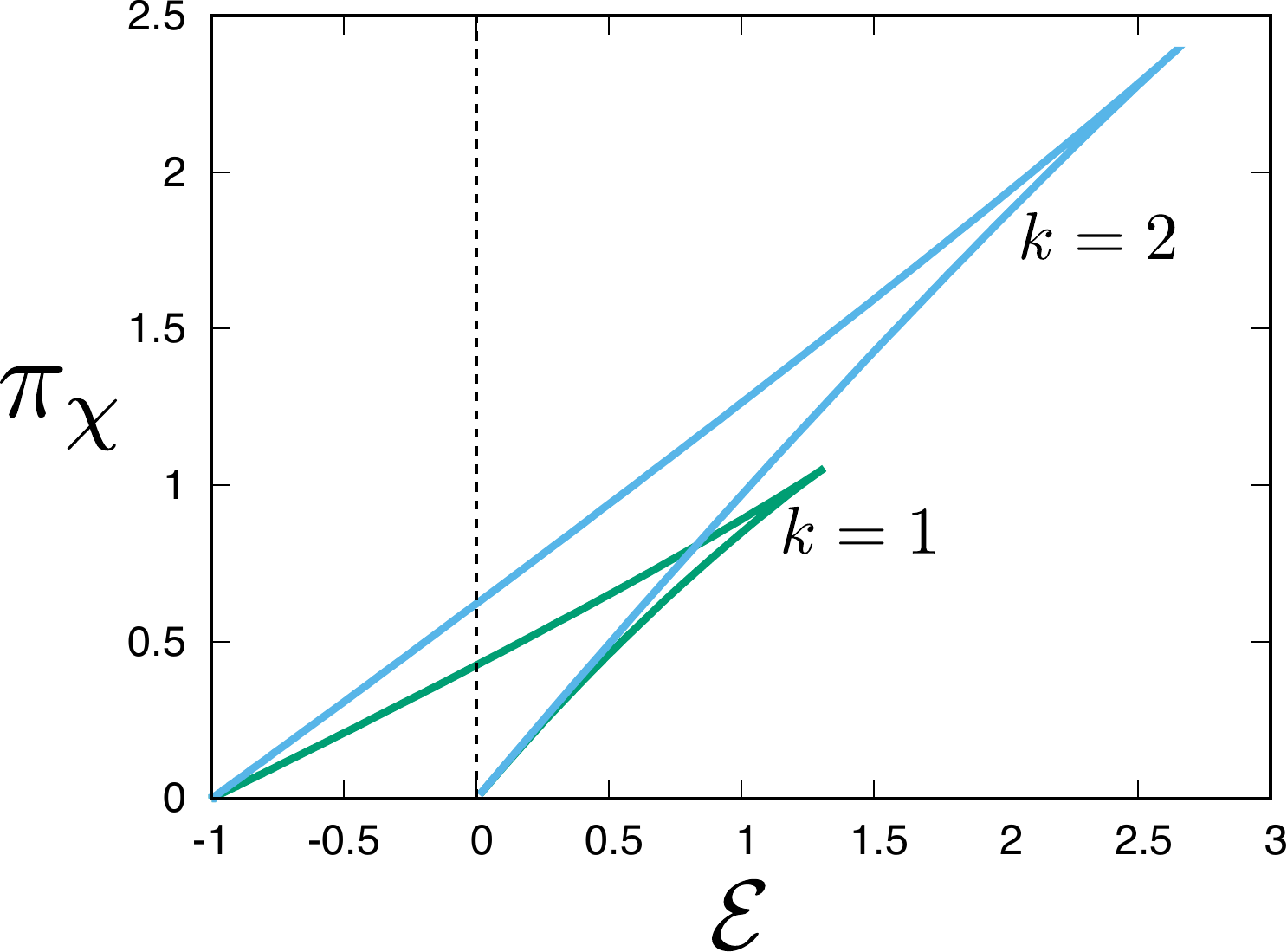}
  }
 \caption{%
Physical quantities of the resonating AdS soliton 
as a function of the energy density.
}
\label{om_P_T_S_k012}
\end{figure}

In the left panel of Fig.~\ref{E_and_z0_vs_a0_k0}, the energy density of the resonating AdS soliton is plotted as a function of $\alpha_0$.
The left edge of the figure, $\alpha_0=0$, corresponds to the AdS soliton.
The energy density is negative when $\alpha_0$ is small but becomes positive as $\alpha_0$ increases. 
It then reaches the maximum value and decreases beyond that point.
For $k=0$, solutions apparently exist until $\alpha_0\to \infty$, where $\mathcal{E}$ approaches zero from above.
On the other hand, for $k=1,2$, the energy density hits zero at a finite value of $\alpha_0$.
As shown in the right panel of Fig.~\ref{E_and_z0_vs_a0_k0}, $z_0 M_{KK} \to \infty$ in the limit $\mathcal{E}\to 0$. This means that the location of the tip $z_0$ diverges, and the spacetime approaches Poincar\'{e} AdS with a compact direction.\footnote{The geometry is highly deformed near the tip $z=z_0$ when $\alpha_0$ is large, but that region, deep in the bulk, is decoupled from the boundary as $z_0 \to \infty$.}
In fact, physical quantities approach those for the pure AdS in that limit as we will see in Fig.~\ref{om_P_T_S_k012}.

In Fig.~\ref{om_P_T_S_k012}, we show the physical quantities of the resonating AdS soliton.
We choose the energy density $\mathcal{E}$ as the horizontal axis.
The left edge, $\mathcal{E}=-1$, corresponds to the AdS soliton limit.
The other endpoints of the curves correspond to the pure AdS limit, where $\mathcal{E}\to 0$.
Because $\mathcal{E}$ takes the maximum value halfway,
physical quantities are multivalued in $\mathcal{E}> 0$.
In particular, two different solutions can be found at $\mathcal{E}=0$: 
resonating AdS soliton and the pure AdS.
As seen in the first panel, the frequency of the resonating AdS soliton satisfies $\omega>k M_{KK}$.
We will show in the next subsection that this inequality results in the non-existence of global timelike Killing vectors.
In the third panel, we find that the tension $\mathcal{T}$ can take negative values for $k=1,2$.
If $\mathcal{T}<0$, negative work is necessary for expanding the radius of the circle with $2\pi/M_{KK}$.
In the figure for $\pi_\chi$ (last panel), cusps can be observed.
Note that $\pi_\chi=0$ for $k=0$, which is omitted in the figure.
For $k \neq 0$, we checked that the first law of thermodynamics $d\mathcal{E} = \omega/(k M_{KK}) d \pi_\chi$ is satisfied within numerical errors.

\subsection{Dynamical spacetime}

Does the resonating AdS soliton spacetime have global timelike Killing vectors?
A general linear combination of the Killing vectors $\{K_1, K_2, \partial_x, \partial_y\}$ can be given by
\begin{equation}
K=c_1 \left(\partial_t + \frac{\omega}{2}\xi_0 \right) + c_2 \left(\partial_\theta-\frac{k}{2}\xi_0 \right)\ ,
\label{generalK}
\end{equation}
where
\begin{equation}
 \xi_0=(x-x_0)\partial_y - (y-y_0)\partial_x\ ,
\end{equation}
and $c_1,\,c_2,\,x_0$ and $y_0$ are arbitrary parameters specifying the linear combination.
The norm of $K$ is given by
\begin{equation}
 K^2=g_{\mu\nu}K^\mu K^\nu
=c_1^2 g_{tt} + 2 c_1c_2 g_{t\theta}+c_2^2g_{\theta\theta}+\frac{(c_1\omega-c_2 k)^2}{4}\xi_0^2\ ,
\label{Ksq}
\end{equation}
where $\xi_0^2=g_{\mu\nu}\xi_0^\mu\xi_0^\nu$.
Near the AdS boundary, $\xi_0^2$ becomes
\begin{equation}
\xi_0^2 \simeq \frac{(x-x_0)^2+(y-y_0)^2}{z^2}\ .
\end{equation}
Thus $\xi_0^2\to \infty $ as $x,y\to \infty$.
This cannot be compensated by $c_1^2 g_{tt} + 2 c_1c_2 g_{t\theta}+c_2^2g_{\theta\theta}$ 
which does not depend on $x$ and $y$.
For this reason, we choose $c_1=k$ and $c_2=\omega$ to eliminate the term of $\xi_0^2$ in Eq.~(\ref{Ksq}) 
and consider a specific combination $K=k\partial_t-\omega\partial_\theta$.
Then, near the AdS boundary, we obtain
\begin{equation}
 K^2\simeq \frac{\omega^2-k^2 M_{KK}^2}{M_{KK}^2 z^2} \ .
\end{equation}
Hence $K$ cannot be timelike unless $\omega<k M_{KK}$.
However, as shown in Fig.~\ref{om_P_T_S_k012}(a), we always have $\omega>k M_{KK}$, and therefore $K^2>0$.
Thus the resonating AdS soliton is a dynamical spacetime.

\subsection{Free energy}
\label{sec:freeE}

The renormalized gravitational Lorentzian action is given by \cite{Balasubramanian:1999re}
\begin{multline}
 I_L=\frac{1}{16\pi G_5} \int_{z\geq \epsilon} d^5 x \sqrt{-\textrm{det}\,g_{\mu\nu}}\left(R+12\right)\\
-\frac{1}{8\pi G_5}\int_{z=\epsilon} d^4x \sqrt{-\gamma}K
-\frac{3}{8\pi G_5}\int_{z=\epsilon} d^4x \sqrt{-\gamma}\left(1+\frac{\mathcal{R}}{12}\right)\ ,
\label{Sos}
\end{multline}
where $\gamma_{ij}$ is the induced metric on the $z=\epsilon$ surface, its determinant is denoted by $\gamma=\textrm{det}\,\gamma_{ij}$, and $\mathcal{R}$ is the Ricci scalar with respect to $\gamma_{ij}$. 
The trace of the extrinsic curvature $K$ is also defined with respect to $\gamma_{ij}$,
\begin{equation}
 K=\gamma_{ij}K^{ij}\ , \qquad K^{ij}= -\frac{1}{2}(\nabla^i n^j + \nabla^j n^i)\ ,
\end{equation}
where $n^i$ is the outward pointing  unit normal vector at $z=\epsilon$.
Note that $R=-20$ from the Einstein equations, and the first term becomes just a volume integral.

For stationary spacetime, the Euclidean on-shell action $I_E$ is obtained by replacing $\int dt \to - \int^{1/T}_0 d\tau$ in the Lorentzian action $I_L$, and the free energy is related to the Euclidean action as $\mathcal{F}=TI_E$.
For general dynamical spacetime, however,
we cannot define the Euclidean action and free energy like stationary spacetime.
Even though the resonating AdS soliton is a dynamical spacetime, the on-shell Lagrangian is fortunately time-independent.
We take advantage of this feature and can define the free energy of the resonating AdS soliton.

For the metric of the resonating AdS soliton~(\ref{coh-1}), we obtain
\begin{equation}
\begin{split}
&\sqrt{-\gamma}=\frac{z_0\sqrt{\beta_\infty}}{2z^4}-\frac{\sqrt{\beta_\infty}(1-c_f-c_\beta)}{4 z_0^3}+\cdots\ ,\\
&\sqrt{-\gamma}K=-\frac{2z_0\sqrt{\beta_\infty}}{z^4}+\frac{\sqrt{\beta_\infty}(1-c_f-c_\beta)}{z_0^3}+\cdots\ ,\\
&\mathcal{R}=\mathcal{O}(z^{10}) \ .
\end{split}
\end{equation}
Substituting these into Eq.(\ref{Sos}) and including the $\mathcal{O}(z^{-4})$ terms into the integrand, we obtain
\begin{equation}
I_L=-\frac{z_0 V_2}{2 G_5} \int dt \left[\int_0^{z_0} \frac{dz}{z^5}
\left(\sqrt{\frac{f\beta}{g}}-\sqrt{\beta_\infty}\right)
-\frac{\sqrt{\beta_\infty}(1+c_f+c_\beta)}{8 z_0^4}
\right]\ ,
\end{equation}
where $V_2=\int dxdy$.
In this expression, the $z$-integral converges at the AdS boundary $z= 0$, and therefore we can set $\epsilon=0$. 
The Euclidean action $I_E$ can be obtained by replacing $\int dt \to -1/T$. Then, the free energy $\mathcal{F}=TI_E$ 
is given by
\begin{equation}
 \mathcal{F}=\frac{z_0 V_2}{2 G_5} \left[\int_0^{z_0} \frac{dz}{z^5}
\left(\sqrt{\frac{f\beta}{g}}-\sqrt{\beta_\infty}\right)
-\frac{\sqrt{\beta_\infty}(1+c_f+c_\beta)}{8 z_0^4}
\right]\ .
\label{Fgeneral}
\end{equation}
For the undeformed AdS soliton, $f(r)=g(r)=\alpha(r)=\beta(r)=1$ and $h(r)=0$, the free energy of the AdS soliton is given by
\begin{equation}
 \mathcal{F}_\textrm{AdS soliton}=-\frac{V_2}{16 G_5 z_0^3}=-\frac{V_2 M_{KK}^3}{128 G_5}\ .
\end{equation}

\begin{figure}
\centering
\includegraphics[scale=0.6]{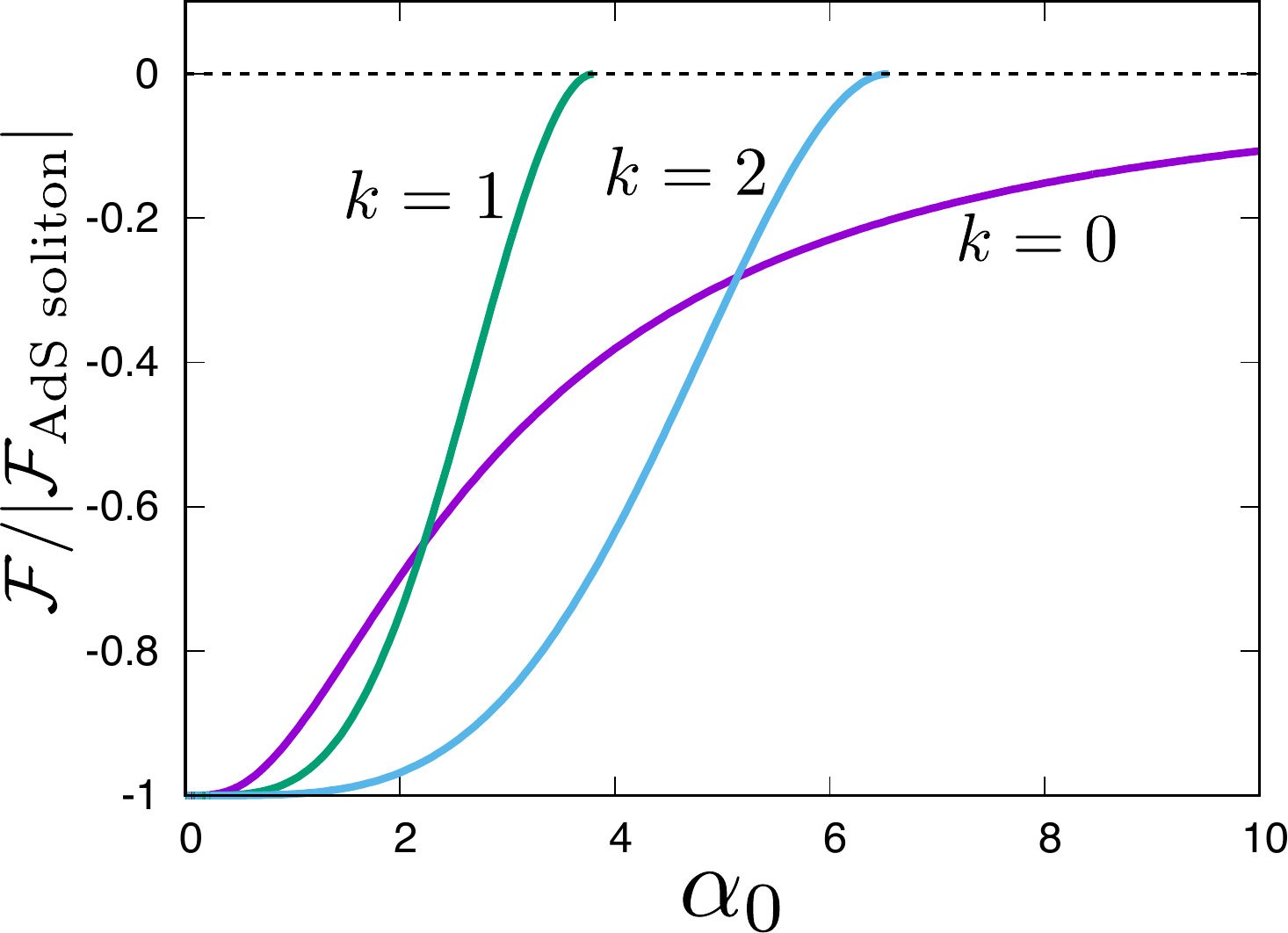}
 \caption{%
Free energy of the resonating AdS soliton as a function of $\alpha_0$. 
}
\label{fig:freeE}
\end{figure}

The free energy of the resonating AdS soliton is shown in Fig.~\ref{fig:freeE} for $k=0,1,2$.
The normalization is given by the absolute value of the free energy of the AdS soliton.
We find that the free energy of the resonating AdS soliton is always bigger than that of the AdS soliton. 
This behavior has already been observed in the perturbative results \cite{Hartnett:2012np}.
We find that it is the case also in fully nonlinear solutions.
The indication of the higher free energy is that the resonating AdS soliton is thermodynamically subdominant in the canonical ensemble. 
For the dual field theory, this is consistent with the no-go theorem for time crystals as the ground state~\cite{Watanabe:2014hea}.

\section{Conclusion}
\label{sec:conclusion}

We constructed the resonating AdS soliton as the nonlinear extension of normal modes of the static AdS soliton. 
We focused on the spin-2 homogeneous perturbation of the AdS soliton, which breaks the isometries \eqref{isometryAdSsoliton} to smaller ones \eqref{isometryG}.
In particular, the time translation of the AdS soliton is broken, and a helical isometry given by $K_1$ is realized.
We introduced the cohomogeneity-1 metric ansatz \eqref{coh-1} and obtained the solutions of the resonating AdS soliton numerically. It was shown that the resonating AdS soliton is a non-stationary dynamical spacetime. The pressure and shear stress behave time periodically, while other thermodynamic quantities are time independent. 

In \cite{Biasi:2019eap}, periodic driving by the boundary source of metric has been studied for the four dimensional AdS soliton, and time dependent solutions have been obtained by solving PDEs.
In the zero amplitude limit of the driving, they become spontaneously time periodic solutions. The resonating AdS soliton is analogous to them. In the present work, however, by going to five dimensions, we were able to use the cohomogeneity-1 metric ansatz \eqref{coh-1} and constructed the time periodic solution without applying boundary driving and dealing with PDEs. 

The energy of the resonating AdS soliton is higher than that of the AdS soliton (see Fig.~\ref{E_and_z0_vs_a0_k0}). The static AdS soliton has been conjectured to be the minimal energy solution among those with the same boundary topology \cite{Horowitz:1998ha,Constable:1999gb},
and the higher energy of the resonating AdS soliton is in accord with this conjecture.\footnote{On a constant $t$ spacelike surface, we obtain $\hat{R} + n(n-1) = \hat{K}_{ij} \hat{K}^{ij} - \hat{K}^2$ from the Hamiltonian constraint, where $n=4$ for the five dimensional bulk, the hats denote the quantities defined on the spacelike surface, $i$ and $j$ run over the spatial coordinates, $\hat{K}_{ij}$ is the extrinsic curvature, and $\hat{K}$ is the trace of $\hat{K}_{ij}$. It is straightforward to check that $\hat{K}=0$ for the metric ansatz (\ref{coh-1}). Therefore, the resonating AdS soliton satisfies $\hat{R} + n(n-1) \ge 0$, and the results for positivity of relative energy in \cite{Barzegar:2019vaj} can be applied.}
The free energy of the resonating AdS soliton is also higher than that of the static AdS soliton as shown in Fig.~\ref{fig:freeE}. Hence, the resonating AdS soliton is thermodynamically subdominant.

For the dual field theory, the non-stationary AdS soliton not being the ground state is consistent with the no-go theorem for time crystals as the ground state \cite{Watanabe:2014hea}. 
The field theory realization of the time periodic solution dual to the resonating AdS soliton remains unclear, but we are sure it would be an excited state. 
It might be interpreted as a homogeneous coherent excitation of glueballs in the confined phase of the Yang-Mills theory and characterized by the time dependence in the pressure and shear stress.

In Fig.~\ref{om_P_T_S_k012}, we observed that the physical quantities are multivalued for a fixed $\mathcal{E}>0$. This behavior typically implies the presence of instability.
On the one hand, it would be interesting to look at linear perturbations of the resonating AdS soliton. In global AdS space, cohomogeneity-1 geons are linearly stable in a large portion of parameter space \cite{Ishii:2020muv}, and the current horizonless solution may share the same property, especially in the branch of small deformation, while the highly deformed branch may have a different behavior and show linear instability instead.

On the other hand, time evolution of the resonating AdS soliton can also be considered directly. At a fixed $\mathcal{E}>0$, a straightforward jump from one of the branches to the other is a possibility, but other dynamics may be involved. Also, the resonating AdS soliton itself is a dynamical spacetime in the first place, and it may develop into a different deformation of the AdS soliton. In \cite{Craps:2015upq,Myers:2017sxr}, quenches and time evolution have been studied for the AdS soliton. Similar computations for the resonating AdS soliton will be interesting.

\acknowledgments
The authors would like to thank Jorge E.~Santos and Benson Way for useful discussions and comments.
The  work of M.~G.~was supported, in part, by the U.S.~Department of Energy grant DE-SC-0012447. M.~G.~acknowledges financial support from the University of Alabama Graduate School, facilitating his visit at Kyoto University and Nihon University, where part of this work was conducted. 
The work of T.~I.~was supported in part by JSPS KAKENHI Grant Number JP18H01214 and JP19K03871.
The work of K.~M.~was supported in part by JSPS KAKENHI Grant Number JP18H01214 and JP20K03976.

\bibliography{bib_reso_ads_soliton}

\end{document}